\documentclass{llncs}
\usepackage{amssymb}
\usepackage{amsmath}
\usepackage{graphicx}
\usepackage{color}
\usepackage[cmtip,arrow]{xy}
\usepackage{pb-diagram,pb-xy}
\def\A{\mathbb{A}} 
\def\Z{\mathbb{Z}}
\def\E{E}
\def\e{\mathrm{e}}
\def\U{\mathrm{U}}
\def\G{G}
\def\Ga{\Gamma}
\def\W{{W}}
\def\X{X}
\def\GaX{\Ga^{\X}}
\def\Partransport{P}
\def\Rtimes{\rtimes}
\def\Times{\times}
\def\State{\sigma}
\def\Stateset{\Sigma} 
\def\StatesetX{\Stateset^{\X}}
\def\Tr{\mathrm{Tr}}
\def\Herm{\dagger}
\def\Hspace{\Psi}
\def\lstep{-} 
\def\rstep{+} 
\def\lprob{p_\lstep} 
\def\rprob{p_\rstep} 
\def\veloc{v} 
\def\PP{\widetilde{P}}
\def\Repbare{\rho}
\newcommand{\Alt}[1]{\mathrm{Alt}\!\left(#1\right)} 
\newcommand{\Cgr}[1]{\Z_{#1}} 
\newcommand{\Dih}[1]{\mathrm{D}_{#1}} 
\newcommand{\Perm}[1]{\mathrm{Sym}\left(#1\right)} 
\newcommand{\Rep}[1]{\Repbare\left(#1\right)} 
\newcommand{\RepH}[1]{\Repbare^\dagger\left(#1\right)} 
\newcommand{\rbra}[1]{\left(#1\right)} 
\newcommand{\set}[1]{\left\{#1\right\}} 
\def\wbrl{\left(}
\def\wbrr{\right)}
\def\wmult{\ast} 
\newcommand{\welem}[2]{\wbrl{#1},~{#2}\wbrr} 
\begin{document}
\title{Discrete Dynamics: Gauge Invariance and Quantization}
\titlerunning{Gauge Invariance and Quantization}
\author{Vladimir V. Kornyak}
\institute{Laboratory of Information Technologies \\
           Joint Institute for Nuclear Research \\
           141980 Dubna, Russia \\
           \email{kornyak@jinr.ru}}
\authorrunning{Vladimir V. Kornyak}
\maketitle
\begin{abstract}
Gauge invariance in discrete dynamical systems and its connection with quantization 
are considered. 
For a complete description of gauge symmetries of a system
we construct explicitly a class of groups unifying in a natural way the \emph{space} 
and \emph{internal} symmetries.
We describe the main features of the gauge principle relevant to the discrete and finite background.
Assuming that continuous phenomena are approximations of more fundamental discrete processes,
we discuss -- with the help of a simple illustration -- relations between such processes and their
continuous approximations. 
We propose an approach to introduce quantum structures in discrete 
systems, based on finite gauge groups. In this approach  quantization can be interpreted
as introduction of gauge connection of a special kind.
We illustrate our approach to quantization by a simple model and suggest generalization 
of this model.
One of the main tools for our study is a program written in C.
\end{abstract}
\section{Introduction}
In 1918 Hermann Weyl -- guided by the concept that the scale of length is arbitrary:
if there is no fundamental length in Nature it does not matter what unit of length is used in
measurements -- conjectured that the scale can be taken in the form $e^{S({x})}$, i.e., 
it may vary from point to point in time and space. This idea failed in application to physics but gave
rise to the concept of \emph{gauge invariance}.
\par
Later --  in 1929, after advent of Quantum Mechanics --
Weyl (and also Vladimir Fock and Fritz London) replaced scale transformations  $e^{S({x})}$
by rotations (phase transformations)  $e^{iS({x})}$ and derived electromagnetism from 
the gauge principle.
\par
In 1954 C.N. Yang and R. Mills extended the gauge principle to non-Abelian symmetries.
Now the gauge principle is recognized as one of the central principles in 
contemporary physics -- in fact, all fundamental physical 
theories are gauge theories (for historical review  see \cite{RaifStrau}).
\par
The lattice gauge theory was introduced by K.G. Wilson in 1974  as 
a practical approach to the problems of strong interactions for which the standard perturbative methods are
inapplicable. This technique -- based on approximation of space, or space-time, by some 
(usually hypercubic) lattice -- was considered as an auxiliary computational method rather 
than a fundamental construction. 
The later mathematical generalizations established relations between lattice gauge theories and such topics
as \emph{topological quantum field theory} (TQFT), \emph{invariants of 3- and 4-manifolds},\emph{ monoidal 
categories}, \emph{Hopf algebras and quantum groups,} \emph{quantum gravity} etc., \cite{Oeckl}.
\par
In view of their origin and applications, the above mentioned lattice gauge theories 
are not entirely discrete constructions. 
They involve continuous ingredients: gauge groups are Lie groups, 
Lagrangians and observables are real or complex functions.
Furthermore, the gauge groups of these theories are groups of internal symmetries and do not involve 
the lattice symmetries. It seems desirable to include the space symmetries into construction of 
gauge group, since: (a) the quantum statistics of particles is characterized by the rules describing 
their behavior under permutations of the points of space; (b) there exist gauge theories that
deduce gravity by interpreting  the space or space-time symmetries as gauge groups.
\par
In this paper we consider more radical version of discrete gauge invariance. 
All our manipulations including 
quantization remain 
within the framework of exact discrete mathematics requiring no more than the ring of algebraic
 integers (and sometimes the quotient field of this ring).
Our study was carried out with the help of a program in C we are developing now.
\section{Discrete Dynamics}
We consider evolution in the discrete time $t\in\Z=\set{\ldots,-1,0,1,\ldots}$.
\par
Let the space  $\X$ be a finite set of points: $\X=\set{x_1,\ldots,x_{N_\X}}$.
This -- primordially amorphous -- set may
possess some structure: some points may be ``closer'' to each other than others.
A mathematical abstraction of such a structure is an \emph{abstract simplicial complex} 
-- a collection of subsets of $\X$ (\emph{simplices}) such that any subset 
of a simplex is also simplex. One-dimensional complexes, i.e., \emph{graphs} 
(or \emph{lattices}), are sufficient to formulate a gauge theory.
The symmetry group of the space $\X$ is the graph automorphism group 
$\G=\mathrm{Aut}\left(\X\right)$.
\par
Table \ref{LatticesX} shows some lattices with their symmetries. 
We use these lattices in our computer experiments. 
In the table $N_{\X}$ and $N_{\E}$ are numbers of points and vertices
in space $\X$; the \emph{trivial}, \emph{symmetric}, \emph{cyclic}, 
\emph{dihedral} and \emph{alternating}
groups are denoted by $1$, $\Perm{n}$, $\Cgr{n}$, $\Dih{2n}$ and $\Alt{n}$, respectively;
the signs $\times$ and $\rtimes$ denote \emph{direct} and \emph{semidirect} products, 
respectively.
Note, that the lattice denoted as \emph{Toric square $n\times{}n$} in the table 
has three times larger symmetry group at $n=4$ than the general case formula predicts%
\footnote{N. Vavilov pointed out to the author that this extra symmetry can be explained by 
$\Cgr{3}$ symmetry of the Dynkin diagram 
$D_4=~~$\raisebox{-0.025\textwidth}{\includegraphics[width=0.06\textwidth]{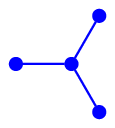}}
 ~~associated with the case $n=4$.}.
\begin{table}[t]
	\centering
		\caption{Examples of discrete spaces}
	  \label{LatticesX}
		\begin{tabular}{c|c|c|c|c}
$\X$&$N_{\X}$&$N_{\E}$&$\G$&$\left|\G\right|$		
\\
\hline
\begin{tabular}{c}
\emph{Atom}\\[-3pt]
\includegraphics[width=0.01\textwidth]{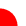}
\end{tabular}
& $1$ & $0$ & $1$ & $1$ 
\\
\hline
\begin{tabular}{c}
\emph{Dimer}\\[-3pt]
\includegraphics[width=0.06\textwidth]{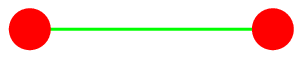}
\end{tabular}
& $2$ & $1$ & $\Perm{2}\cong\Cgr{2}$ & $2$ 
\\
\hline
\begin{tabular}{c}
\emph{Triangle}\\[-1pt]
\includegraphics[width=0.06\textwidth]{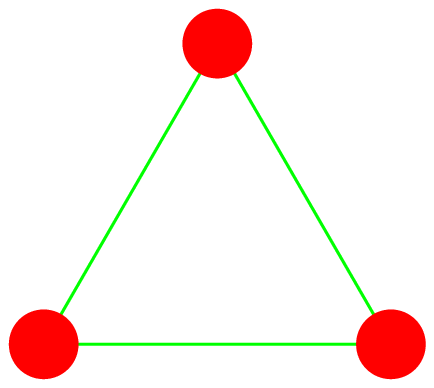}
\end{tabular}
& $3$ & $3$ & $\Perm{3}\cong\Dih{6}$ & $6$
\\
$n$-\emph{vertex polygon}& $n$ & $n$ & $\Dih{2n}$ & $2n$
\\
\hline
\begin{tabular}{c}
\emph{Tetrahedron}\\[-3pt]
\includegraphics[width=0.075\textwidth]{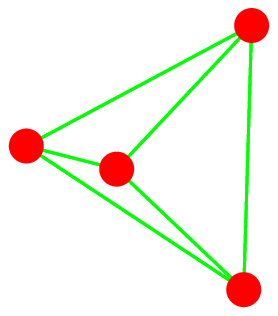}
\end{tabular}
& $4$ & $6$ & $\Perm{4}$ & $24$
\\
\hline
\begin{tabular}{c}
\emph{Octahedron}\\[-3pt]
\includegraphics[width=0.095\textwidth]{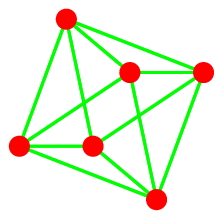}
\end{tabular}
& $6$ & $12$ & $\Cgr{2}\times\Perm{4}$ & $48$
\\
\hline
\begin{tabular}{c}
\emph{Hexahedron}\\[-2pt]
\includegraphics[width=0.095\textwidth]{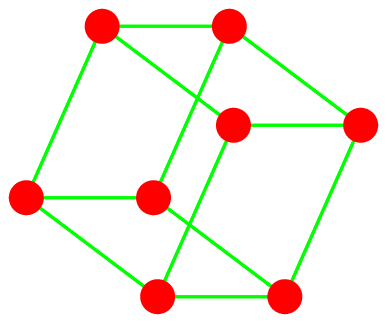}
\end{tabular}
& $8$ & $12$ & $\Cgr{2}\times\Perm{4}$ & $48$
\\
\hline
\emph{Toric square $n{}\times{}n, n{}\neq{}4$}& $n^2$ & $2n^2$ & 
$\left(\Cgr{n}\times\Cgr{n}\right)\rtimes\Dih{8}$
 & $8n^2$
\\
\hspace*{-35pt}
\emph{$n=4$}
\raisebox{-0.04\textwidth}{\includegraphics[width=0.1\textwidth]{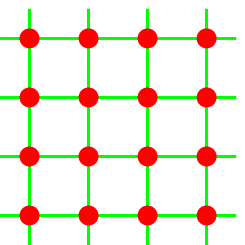}}
& $16$ & $32$ 
& 
$\left(\left(\left(\left(\Cgr{2}{}\times\Dih{8}\right){} 
\rtimes\Cgr{2}\right){}\rtimes%
{\color{red}\Cgr{3}}%
\right){}\rtimes\Cgr{2}\right)\rtimes\Cgr{2}$
& 
$384$
\\
\hline
\begin{tabular}{c}
\emph{Icosahedron}\\[-2pt]
\includegraphics[width=0.1\textwidth]{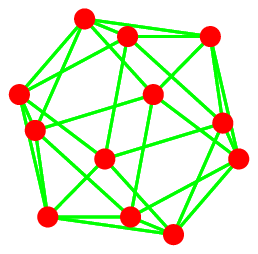}
\end{tabular}
& $12$ & $30$ & $\Cgr{2}\times\Alt{5}$ & $120$
\\
\hline
\begin{tabular}{c}
\emph{Dodecahedron}\\ 
\includegraphics[width=0.12\textwidth]{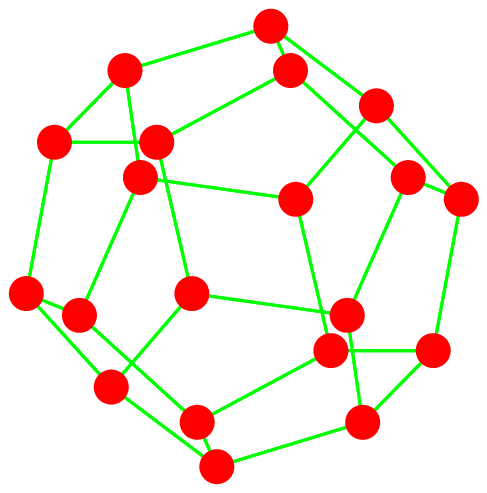}
\end{tabular}
& $20$ & $30$ & $\Cgr{2}\times\Alt{5}$ & $120$
\\
\hline
\begin{tabular}{c}
\emph{Fullerene $C_{60}$}\\ 
\includegraphics[width=0.16\textwidth]{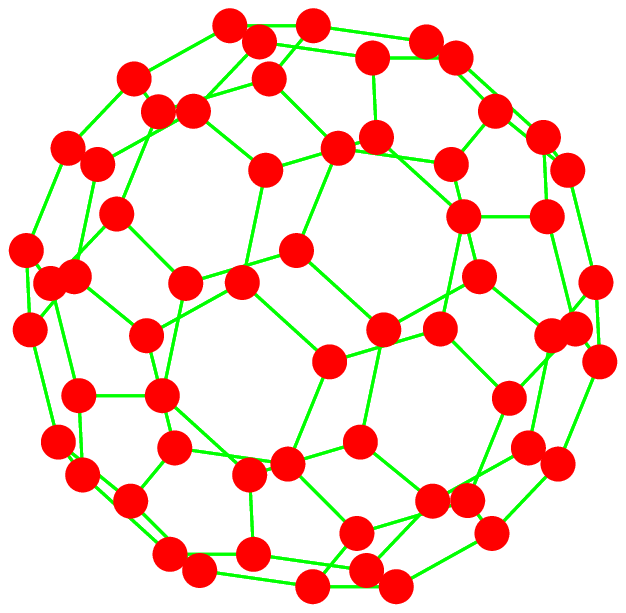}
\end{tabular}
& $60$ & $90$ & $\Cgr{2}\times\Alt{5}$ & $120$
\\
\hline
\begin{tabular}{c}
\emph{Toric graphene $n\times m$}\\ 
\includegraphics[width=0.15\textwidth]{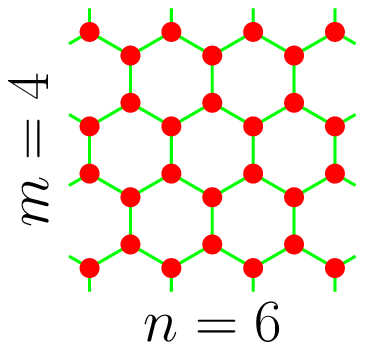}
\end{tabular}
& $nm$ & ${}\frac{\textstyle{3nm}}{\textstyle{2}}{}$ & $\Dih{n}\times\Dih{2m}$ & $2nm{}$
\\
\hline
		\end{tabular}
\end{table}
\par
Let each point $x\in\X$ take values in some finite \emph{set of local states}
$\Stateset=\set{\State_1,\ldots,\State_{N_{\Stateset}}}$ possessing
 some symmetry group $\Ga\leq\Perm{\Stateset}.$ 
 Such groups are analogs of the ``groups of internal
symmetries'' responsible for interactions in 
\clearpage  
\noindent
physical gauge theories. 
The state of a system as a whole is a function $\State(x)\in\StatesetX.$
\par
\emph{Dynamics} of the system is determined by some \emph{evolution rule} connecting 
the current state of the system $\State_{t}(x)$ with its prehistory 
$\State_{t-1}(x),~\State_{t-2}(x),\ldots\enspace.$\\
A typical form of evolution rule is \emph{evolution relation}:
\begin{equation}
R\left(\State_t\left(x\right);~\State_{t-1}\left(x\right),
~\State_{t-2}\left(x\right),\ldots\right)\subseteq\StatesetX
\times\StatesetX\times\cdots\enspace.
\label{rellaw} 
\end{equation}
Most commonly used in applications and convenient for study are \emph{deterministic}
(or \emph{causal}) \emph{dynamical systems}. The current state of deterministic 
system is uniquely determined by its prehistory, i.e., relations like (\ref{rellaw}) 
are \emph{functional} and can be written in the form
\begin{equation*}
\State_t\left(x\right) = F\left(\State_{t-1}\left(x\right), 
\State_{t-2}\left(x\right),\ldots\right)\enspace.\label{detlaw} 
\end{equation*}
There are two important special types of \emph{non-deterministic} dynamical systems: 
\begin{itemize}
\item  
\emph{lattice models in statistical mechanics} -- special instances of Markov chains; 
\item 
\emph{discrete quantum systems}
obtained from classical systems by identification of their states with basis elements of 
complex Hilbert spaces.
\end{itemize}
For these systems transition from one state to any other is possible with some probability
controlled by additional structures: real (for Markov chains) or complex (for quantum systems)
weights assigned to state transitions. In this paper we restrict our attention to
the case of discrete quantum systems.
\section{Unification of Space and Internal Symmetries}
Having the groups $\G$ and $\Ga$ acting on $\X$ and $\Stateset$, respectively, we can 
combine them into a single group $\W$ which acts on the states $\StatesetX$ of the whole system.
The group $\W$ can be identified, as a set, 
with the \emph{Cartesian product} $\GaX\otimes\G$, where $\GaX$ is the set of $\Ga$-valued
functions on $\X.$ That is, every element $u\in\W$ can be 
represented in the form $u = \welem{\alpha(x)}{a},$ where $\alpha(x)\in\GaX$ and $a\in\G.$
\par
In physics, it is usually assumed that the space and internal symmetries are independent, i.e.,
$\W$ is the \emph{direct product} $\Ga^\X\Times\G$ with action%
\footnote{We write group actions \emph{on the right}. 
This, more intuitive, convention is adopted in both GAP and MAGMA -- the
most widespread computer algebra systems with advanced facilities for computational group theory.}
on $\StatesetX$ and multiplication rule:
\begin{eqnarray}
\sigma(x)\welem{\alpha\left(x\right)}{a}&=&\sigma\left(x\right)\alpha\left(x\right)
\text{~~~~~~~~~~~\emph{action}}\enspace,\nonumber \\
	\welem{\alpha\left(x\right)}{a}\wmult\welem{\beta\left(x\right)}{b}&=&
	\welem{\alpha\left(x\right)\beta\left(x\right)}{ab}\label{wmultdir}
	\text{~~~~\emph{multiplication}}\enspace. 
\end{eqnarray}
Another standard construction is the \emph{wreath product} $\Ga\wr_{\X}\G$ 
	having a structure of the semidirect product $\GaX\Rtimes\G$ with action and multiplication
\begin{eqnarray}
\sigma(x)\welem{\alpha\left(x\right)}{a}&=&\sigma\left(xa^{-1}\right)\alpha\left(xa^{-1}\right)\enspace,
\nonumber \\
\welem{\alpha\left(x\right)}{a}\wmult\welem{\beta\left(x\right)}{b}&=&
\welem{\alpha\left(x\right)\beta\left(xa\right)}{ab}\label{wmultwr}\enspace. 
\end{eqnarray}
These examples are generalized by the following
\emph{\textbf{Statement:}}\\
{ There are equivalence classes of \emph{split group extensions}
$1\rightarrow\GaX\rightarrow\W\rightarrow\G\rightarrow1$ 
determined by \emph{antihomomorphisms} $\mu: \G \rightarrow \G$.
The equivalence is described by \emph{arbitrary} function $\kappa: \G \rightarrow \G.$
The explicit formulas for main group  operations --- \textbf{\emph{action}} on $\StatesetX$, \textbf{\emph{multiplication}}
and  \textbf{\emph{inversion}} --- are
\begin{eqnarray}
	\sigma(x)\welem{\alpha\left(x\right)}{a}&=&\sigma\left(x\mu(a)\right)\alpha\left(x\kappa(a)\right)
	\enspace,
	\label{wact}\\[3pt]
	\welem{\alpha\left(x\right)}{a}\wmult\welem{\beta\left(x\right)}{b}&=&
	\welem{\alpha\left(x\kappa(ab)^{-1}\mu(b)\kappa(a)\right)\beta\left(x\kappa(ab)^{-1}\kappa(b)\right)}
	{ab}\enspace, \label{wmult}\\
	\welem{\alpha(x)}{a}^{-1}
		&=&\welem{\alpha\left(x\kappa\left(a^{-1}\right)^{-1}\mu(a)^{-1}\kappa(a)\right)^{-1}}{a^{-1}}
		\enspace.
	\label{winv}
\end{eqnarray}
}
This statement follows from the general description of the structure of split 
extensions 
of a group $G_1$ by a group $G_0$:
all such extensions are determined by the homomorphisms from $G_1$ to $\mathrm{Aut}\left(G_0\right)$ 
(see, e.g., \cite{Kirillov}, p. 18). Specializing this description to the case when $G_0$ is 
the set of $\Ga$-valued function on $\X$ and $G_1$ acts on arguments  of these functions we obtain our statement.
The \emph{equivalence} of extensions with the same antihomomorfism $\mu$ but with different functions $\kappa$
is expressed by the commutative diagram
\begin{equation}
\begin{diagram}
\node{1}
\arrow[1]{e}
\node{\GaX}
\arrow[1]{e}
\arrow[1]{s,=}
\node{\W}
\arrow[1]{e}
\arrow[1]{s,l}{K}
\node{\G}
\arrow[1]{e}
\arrow[1]{s,=}
\node{1}\\
\node{1}
\arrow[1]{e}
\node{\GaX}
\arrow[1]{e}
\node{~\W'}
\arrow[1]{e}
\node{\G}
\arrow[1]{e}
\node{1}
\end{diagram}\enspace,
\end{equation}
where the mapping $K$ takes the form
$
K: \welem{\alpha(x)}{a}\mapsto\welem{\alpha\left(x\kappa(a)\right)}{a}\enspace.
$
\par
Note that the standard direct (\ref{wmultdir})  and wreath (\ref{wmultwr}) 
products are obtained from this general construction by choosing 
$\left(\mu(a)=1, \kappa(a)=1\right)$
 and $\left(\mu(a)=a^{-1},\right.$ $\left.\kappa(a)=a^{-1}\right)$,
 respectively.
\par
In our C program the group $\W$ is specified by two groups $\G$ and $\Ga$ and 
two functions $\mu(a)$ and $\kappa(a)$ implemented as arrays. It is convenient in computations 
 to use the
following specialization: $\mu(a)=a^{-m}$ and $\kappa(a)=a^k$.
For such a choice formulas (\ref{wact}-\ref{winv}) take the form 
\begin{eqnarray}
	\sigma(x)\welem{\alpha\left(x\right)}{a}&=&\sigma\left(xa^{-m}\right)\alpha\left(xa^k\right)
	\enspace,
	\label{wactspec}\\[3pt]
	\welem{\alpha\left(x\right)}{a}\wmult\welem{\beta\left(x\right)}{b}&=&
	\welem{\alpha\left(x(ab)^{-k-m}a^{k+m}\right)\beta\left(x(ab)^{-k}b^k\right)}
	{ab}\enspace, \label{wmultspec}\\
	\welem{\alpha(x)}{a}^{-1}
		&=&\welem{\alpha\left(xa^{2k+m}\right)^{-1}}{a^{-1}}
		\enspace.
	\label{winvspec}
\end{eqnarray}
Here $k$ is \emph{arbitrary} integer, $m=0$ (\emph{direct} product) or $m=1$ (\emph{wreath} product).
\section{Discrete Gauge Principle}
In fact, the gauge principle expresses the very general idea that any observable data can be presented
in different ``frames'' at different points of space and time, and there should be some way to compare
these data.
At the set-theoretic level, i.e., in the form suitable for both discrete and continuous cases, 
the main concepts of the gauge principle can be reduced to the following elements
\def\vspa{}
\begin{itemize}
\vspa
	\item a set $\X$, space or space-time;
\vspa
	\item a set $\Stateset$, local states;
\vspa
	\item the set $\StatesetX$ of $\Stateset$-valued functions on $\X$, the set of states of \emph{dynamical system};
\vspa \vspa \vspa \vspa \vspa
	\item a group $\W\leq\Perm{\StatesetX}$ acting on $\StatesetX$, \emph{symmetries of the system};
\vspa \vspa \vspa
	\item identification of data describing dynamical system with states from $\StatesetX$ makes sense only modulo
	symmetries from $\W$;
\vspa
	\item having no \textit{a priori} connection between data from $\StatesetX$ at different points $x$ and $y$ 
	in time and space we impose this \emph{connection} (or \emph{parallel transport})
	explicitly as $\W$-valued functions on edges of
	 \emph{abstract} graph:
	 $$
	   \Partransport(x,y)\in\W,~\varsigma(y) = \State(x)\Partransport(x,y)\enspace;
	 $$
	 connection $\Partransport(x,y)$ has obvious property $\Partransport(y,x)=\Partransport(x,y)^{-1};$
\vspa
	\item connection $\Partransport(x,y)$ is called \emph{trivial} if it can be expressed 
	in terms of a function on \emph{vertices} of the graph: 
	$\Partransport(x,y)=p(x)^{-1}p(y),~p(x),p(y)\in\W;$
\vspa
	\item invariance  with respect to gauge symmetries 
	depending on time or space $u(x), u(y)\in\W$  
	leads to transformation rule for connection
\begin{equation}
		\Partransport(x,y) \rightarrow u(x)^{-1}\Partransport(x,y)u(y);\label{ginv}
\end{equation}
\vspa
  \item the \emph{curvature} of connection $\Partransport(x,y)$ is defined as the conjugacy class of 
  the \emph{holonomy} along a cycle of a graph: 
    $$
    \Partransport(x_1,x_2,\ldots,x_k)=\Partransport(x_1,x_2)\Partransport(x_2,x_3)\cdots \Partransport(x_k,x_1)
    $$
   (the conjugacy 
   means 
   $\Partransport'(x_1,\ldots,x_k)\sim u^{-1}\Partransport(x_1,\ldots,x_k)u$ for any
   $u\in\W$);\\
   the curvature of trivial connection is obviously trivial:
   $\widetilde{\Partransport}(x_1,\ldots,x_k)\equiv1;$
\vspa
   \item the gauge principle does not tell us anything about the evolution of the connection
   itself, so gauge invariant relation describing dynamics of connection 
   (\emph{gauge field}) should be added.
\end{itemize}
Let us give two illustrations of how these concepts work in continuous case.
\vspa
\subsubsection*{Electrodynamics. Abelian prototype of all gauge theories.}
Here the set $\X$ is 4-dimensional Minkowski space with points $x=\left(x^\mu\right)$ 
and the set of states is Hilbert space of complex scalar (Schr\"{o}dinger equation)
or spinor (Dirac equation) fields $\psi(x).$ The symmetry group of the Lagrangians and physical 
observables is $\W=\U(1).$ The elements of $\U(1)$ can be represented
as $e^{-i\alpha}.$ 
\par
Let us make these elements dependent on space-time and consider the 
parallel transport for two closely situated space-time points: 
$$P(x, x+\Delta x) = e^{-i\rho(x, x+\Delta x)}\enspace.$$
Specializing transformation rule (\ref{ginv}) to this particular case
$$
P'(x, x+\Delta x) = e^{i\alpha(x)}P(x, x+\Delta x)e^{-i\alpha(x+\Delta x)}\enspace,
$$ 
substituting approximations
$$P(x, x+\Delta x) = e^{-i\rho(x, x+\Delta x)}\approx1-i{A}(x)\Delta x\enspace,$$
$$P'(x, x+\Delta x) = e^{-i\rho(x, x+\Delta x)}\approx1-i{A'}(x)\Delta x\enspace,$$
$$e^{-i\alpha(x+\Delta x)}\approx e^{-i\alpha(x)}\left(1-i\nabla\alpha(x)\Delta x\right)\enspace,$$
and taking into account commutativity of $\W=\U(1)$ we obtain
\begin{equation}
{A'}(x)={A}(x)+\nabla\alpha(x)\enspace~~
\mbox{or, in components,}\enspace~~
{A'_\mu}(x)={A_\mu}(x)
+\frac{\textstyle{\partial\alpha(x)}}{\textstyle{\partial x^\mu}}\enspace.\label{grad}	
\end{equation}
The 1-form $A$ taking values in the Lie algebra of $\U(1)$ and its differential
$F=\left(F_{\mu\nu}\right)=\mathrm{d}A$ are identified
with the electromagnetic \emph{vector potential} and  \emph{field strength},
respectively. To provide the gauge invariance of the equations for field $\psi(x)$ we should replace
partial  by covariant derivatives
$$
\partial_\mu \rightarrow D_\mu=\partial_\mu-iA_\mu(x)
$$
in those equations.
\par 
Finally, evolution equations  for the gauge field $A(x)$ should be added. In the case of 
electromagnetics these are Maxwell's equations:
\begin{eqnarray}
\mathrm{d}F&=&0\hspace*{20pt}\text{\emph{first pair}}\label{m1st}\\	
\mathrm{d}\star F&=&0\hspace*{20pt}\text{\emph{second pair}}\label{m2nd}.	
\end{eqnarray}
Here $\star$ is the \emph{Hodge conjugation} (\emph{Hodge star operator}).
Note that equation (\ref{m2nd}) corresponds to \emph{vacuum Maxwell's equations}. In the presence of the  
\emph{current} $J~$ the \emph{second pair} takes the form $~~\star\mathrm{d}\star F=J.$ 
Note also that the \emph{first pair}
is essentially \emph{a priori} statement, it reflects simply the fact that $F$, by definition, is the 
differential of an exterior form.
\subsubsection*{Non-Abelian gauge theories in continuous space-time.} 
Only minor modifications are needed for the case of non-Abelian Lie group $\W.$
Again expansion of the $\W$-valued parallel transport for two close space-time points $x$ 
and $x+\Delta x$ with taking into account
that $P(x,x)=1$ leads to introducing of a Lie algebra valued 1-form $A=\left(A_\mu\right):$
$$
P(x,x+\Delta x)\approx1+A_\mu(x)\Delta x^\mu\enspace.
$$
Infinitesimal manipulations with formula (\ref{ginv})
$$
u(x)^{-1}P(x,x+\Delta x)u(x+\Delta x)~ \longrightarrow~
u(x)^{-1}\left(1+A_\mu(x)\Delta x^\mu\right)
\left(u(x)+\frac{\textstyle{\partial u(x)}}{\textstyle{\partial x^\mu}}\Delta x^\mu\right)
$$
lead to the following transformation rule
\begin{equation}
{A'_\mu}(x)=u(x)^{-1}{A_\mu}(x)u(x)
+u(x)^{-1}\frac{\textstyle{\partial u(x)}}{\textstyle{\partial x^\mu}}
\enspace.\label{nagrad}	
\end{equation}
The curvature 2-form
$$
F = dA+\left[A\wedge A\right]
$$
is interpreted as \emph{physical strength field}. In particular, the \emph{trivial} connection
 $$\widetilde{A}_\mu(x)=u_0(x)^{-1}\frac{\textstyle{\partial u_0(x)}}{\textstyle{\partial x^\mu}}$$ 
is \emph{flat}, i.e., its curvature $F=0.$ 
\par
There are different approaches to construct dynamical equations for gauge fields
\cite{Oeckl}. The most important example is \emph{Yang-Mills theory} based on the Lagrangian
$$
L_{YM}=\mathrm{Tr}\left[F\wedge\star F\right]\enspace.
$$
The Yang-Mills equations of motion read 
\begin{eqnarray}
\mathrm{d}F+\left[A\wedge F\right]&=&0\enspace,\label{Bianci}\\
\mathrm{d}\star F+\left[A\wedge\star F\right]&=&0\enspace.
\end{eqnarray}
Here again equation (\ref{Bianci}) is  \emph{a priori} statement called 
\emph{Bianci identity}. 
Note that Maxwell's equations are a special case of Yang-Mills equations. 
\par
It is instructive to see what the Yang-Mills Lagrangian looks like in the discrete approximation.
Replacing the Minkowski space $\X$ by a hypercubic lattice one can see that the discrete
version of $L_{YM}$ is proportional to $\sum_f\sigma\left(\gamma_f\right)$,
where the summation is over all faces of a hypercubic constituent of the lattice;
$$\sigma = 2\dim{}U-\left(\chi_U+\chi_{U^{\Herm}}\right);$$ $\chi_U$ and $\chi_{U^{\Herm}}$ are 
characters of the fundamental representation $U$ of the gauge group and its dual representation,
respectively; $\gamma_f$ is the gauge group holonomy around the face $f$.
\par
The Yang-Mills theory 
uses  Hodge operation converting $k$-forms to $(n-k)$-forms
in $n$-dimensional space \emph{with metric} $g_{\mu\nu}$.
In topological applications so-called \emph{BF theory} plays
an important role since it does not involve a metric. 
In this theory, an additional dynamical field $B$
is introduced. The Lie algebra valued $(n-2)$-form $B$ and the $2$-form
$F$ are combined into the  Lagrangian $L_{BF}=\mathrm{Tr}\left[B\wedge F\right]\enspace.$
\section{Quantization Based on Finite Group}
\emph{Quantization} is a procedure for recovering a more fundamental quantum theory
from its classical approximation.
Both Lagrangian and Hamiltonian formulations of classical mechanics 
are based on the \emph{principle of least action} which
 looks a bit mysterious: a particle moving from one point to another ``knows'' 
in advance where it is going to arrive. 
Feynman's path integral quantization \cite{Feynman} eliminates this apparent teleology in a 
quite natural way:
classical trajectories correspond to the dominating (in the path integral) part of all possible trajectories.
\par
Of course, recovering a theory from its approximation can not be performed uniquely.
Moreover,  discrepancies between a theory and its approximation may be essential.
To illustrate, let us compare a simple discrete process with its approximation by continuous physical law.
\subsection{Heat Equation from Bernoulli Trials}
Let us consider a sequence of Bernoulli trials.
The probability of a separate sequence is described by the \emph{binomial distribution}
\begin{equation}
P\left(n_\lstep,n_\rstep\right)=\frac{\left(n_\lstep+n_\rstep\right)!}{n_\lstep!n_\rstep!}
\lprob^{n_\lstep}\rprob^{n_\rstep}\enspace.\label{bindistr}
\end{equation}
Here 
$\set{\lstep,~\rstep}$ are possible outcomes of a single trial; 
$\lprob,~\rprob$ are probabilities ($\lprob+\rprob=1$)
and $n_{\lstep},~n_{\rstep}$ are numbers of the outcomes.
\par
Applying~
Stirling's approximation~ to (\ref{bindistr}) and introducing new variables 
$x = n_\rstep-n_\lstep,~ t = n_\lstep+n_\rstep,~ \veloc=\lprob-\rprob$
--- let us call them ``space'', ``time'' and ``velocity'', respectively ---  we obtain
\begin{equation}
	P\left(x,t\right)\approx \PP\left(x,t\right)
	=\frac{1}{\sqrt{1-\veloc^2}}\sqrt{\frac{2}{\pi{}t}}
	\exp\left\{-\frac{1}{2t}\left(\frac{x-\veloc t}{\sqrt{1-\veloc^2}}\right)^2\right\}\enspace.
	\label{Pxt}
\end{equation}
This is the \emph{fundamental solution} of the \emph{heat} (also known as
 \emph{diffusion} or \emph{Fokker--Planck}) \emph{equation}:
\begin{equation}
	\frac{\partial \PP\left(x,t\right)}{\partial t}
	+\veloc\frac{\partial \PP\left(x,t\right)}{\partial x} 
	=\frac{\left(1-v^2\right)}{2}
	\frac{\partial^2 \PP\left(x,t\right)}{\partial x^2}\enspace.
	\label{heq}
\end{equation}
\par
Note that expression (\ref{Pxt}) contains ``relativistic'' fragment 
$\frac{\textstyle{x-\veloc t}}{\textstyle{\sqrt{1-\veloc^2}}}$ due to 
 the velocity limits $-1\le\veloc\le1$ in our model. 
 Note also that at $\left|\veloc\right|=1$ equation (\ref{heq})
reduces to the \emph{wave equation}
\begin{equation}
	\frac{\partial \PP\left(x,t\right)}{\partial t}\pm\frac{\partial 
	\PP\left(x,t\right)}{\partial x} 
	=0\enspace.
	\label{weq}
\end{equation}
\par
Now let us set a problem as is typical in mechanics: find extremal trajectories connecting 
two fixed points $(0,0)$ and $(X,T)$. We adopt here the search of trajectories with maximum 
probability as a version of the ``least action principle''. 
The probability of trajectory passing through some
intermediate point $(x,t)$ is the following \emph{conditional probability}
\begin{eqnarray}
			P_{(0,0)\rightarrow(x,t)\rightarrow(X,T)}&=&\frac{P(x,t)P(X-x,T-t)}{P(X,T)}\nonumber\\
				&=&\frac{t!(T-t)!\left(\frac{T-X}{2}\right)!\left(\frac{T+X}{2}\right)!}
			{\left(\frac{t-x}{2}\right)!\left(\frac{t+x}{2}\right)!
			\left(\frac{T-t}{2}-\frac{X-x}{2}\right)!\left(\frac{T-t}{2}+\frac{X-x}{2}\right)!T!}\enspace.
			\label{exactp}
\end{eqnarray}
The conditional probability computed for approximation (\ref{Pxt}) takes the form
\begin{equation}
		\PP_{(0,0)\rightarrow(x,t)\rightarrow(X,T)}=\frac{T}{\sqrt{\frac{\pi}{2}(1-\veloc^2)tT(T-t)}}
	\exp\left\{-\frac{\left(Xt-xT\right)^2}{2(1-\veloc^2)tT(T-t)}\right\}.
	\label{approxp}
\end{equation}
One can see essential differences between (\ref{exactp}) and (\ref{approxp}):
\begin{itemize}
	\item exact probabilities (\ref{exactp}) do \emph{not depend} on the velocity $\veloc$
	(or on the probabilities $\lprob,~\rprob$ of a single trial), whereas (\ref{approxp})
	contains \emph{artificial dependence}, 
	\item it is easy to check that expression (\ref{exactp}) allows 
	\emph{many} trajectories
	 with the \emph{same maximum probability}, whereas extremals of (\ref{approxp}) are
	  \emph{deterministic trajectories}, namely, straight lines ~$x = \frac{X}{T}t.$
\end{itemize}
These artifacts show that an important guiding principle of quantization --- correspondence with
classical limit --- may not be quite reliable.
\subsection{Gauge Connection and Quantization} 
\begin{figure}[!h]
\centering
\includegraphics[width=0.85\textwidth]{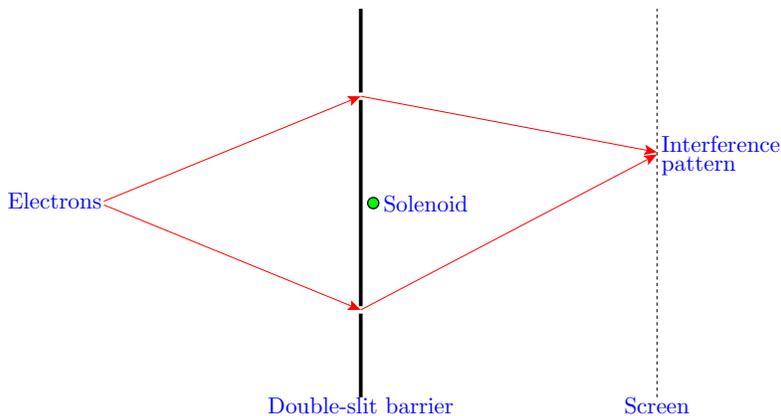}
\caption{Aharonov--Bohm effect. Magnetic flux is confined within the perfectly shielded solenoid;
interference pattern is shifted in spite 
of absence of electromagnetic forces acting on the particles.}
	\label{Aharonov-Bohm}
\end{figure}

The Aharonov--Bohm effect (Fig. \ref{Aharonov-Bohm}) is one of the most striking 
illustrations of interplay between quantum behavior and gauge connection. 
Charged particles moving through 
the region containing perfectly shielded thin solenoid 
produce different interference patterns on a screen depending on whether the solenoid is turned on or off. 
There is no electromagnetic force acting on the particles, but working solenoid produces 
$\U(1)$-connection adding or subtracting phases of the particles and, thus, changing the  interference pattern. 
\par
In the discrete time Feynman's path amplitude decomposes into 
the product of elements of the group $\U(1)$ (or, more precisely, 
elements of the fundamental representation of  $\U(1)$):
\begin{equation}
	 A_{\U(1)}=\exp\left(iS\right)=\exp\left(i\int Ldt \right)\longrightarrow
	  \e^{\textstyle{iL_{0,1}}}\ldots\e^{\textstyle{iL_{t-1,t}}}
	  \ldots\e^{\textstyle{iL_{T-1,T}}}\enspace.\label{FPA}
\end{equation}
By the notation $L_{t-1,t}$  we emphasize that the Lagrangian is in fact a function
defined on pairs of points (graph edges) --- this is compatible with physics 
where the typical Lagrangians are determined by the \emph{first order} derivatives. 
Thus, the expression $\Partransport(t-1,t)=e^{\textstyle{iL_{t-1,t}}}\in\U(1)$ 
can be interpreted as $\U(1)$-parallel transport.
A natural generalization of this is
to suppose that: 
\begin{itemize}
	\item the group $\U(1)$ can be replaced by some other group $\Ga$, 
	\item unitary representation $\Rep{\Ga}$ may have
dimensionality different from 1.
\end{itemize}
\par
We can introduce  quantum mechanical description of a discrete system interpreting 
states $\State\in\Stateset$ as basis elements of a Hilbert space $\Hspace$.
This allows to describe statistics of observations of $\State^,s$
in terms of the \emph{inner product} in $\Hspace$.
\par
Now let us replace expression (\ref{FPA}) for Feynman's path amplitude  by the following 
parallel transport along the path
\begin{equation*}
	 A_{\Rep{\Ga}}=\Rep{\alpha_{0,1}}\ldots\Rep{\alpha_{t-1,t}}
	  \ldots\Rep{\alpha_{T-1,T}}\enspace.\label{Gammaampl}
\end{equation*}
Here $\alpha_{t-1,t}$ are elements of a \emph{finite group}
$\Ga$ -- we shall call $\Ga$ \emph{quantizing group} -- and $\Repbare$ is 
an unitary representation of  $\Ga$ on the space $\Hspace$.
\par
Let us recall main properties of linear representations of finite groups \cite{Serre}.
\begin{itemize}
\item First of all, any linear representation of finite group is equivalent to unitary.
	\item Any unitary representation $\Repbare$ is determined 
	uniquely (up to isomorphism) by its \emph{character} defined as
	 $~~\chi_{\Repbare}(\alpha)=\Tr\Repbare(\alpha),~\alpha\in\Ga.$
\item All values of $\chi_{\Repbare}$ and eigenvalues of $\Repbare$ are elements of 
the ring  $\A$ of \emph{algebraic integers}, moreover the eigenvalues are \emph{roots of unity.}
Recall that the ring  $\A$ consists of the roots of 
 \emph{monic} polynomials with integer coefficients \cite{Kirillov}.
  \item If all different irreducible 
   representations of $\Ga$ are
    $\Repbare_1,\cdots,\Repbare_i,\cdots,\Repbare_h$ 
  and $d_i=\dim\Repbare_i$,
   $M=\left|\Ga\right|$ then 
   $${\sum\limits_{i-1}^h{}d_i^2=M} 
   \text{~~and any~~} {d_i} \text{~~divides~~} 
   {M}:~~ {d_i\mid{}M}.$$
\item Any function $\varphi(\alpha)$ depending only on conjugacy classes of $\Ga$, i.e.,\\
 $\varphi\left(\beta^{-1}\alpha\beta\right)=\varphi\left(\alpha\right)$, 
  is linear combination of characters $\chi_{\Repbare_1},\cdots,\chi_{\Repbare_h}$.\\
  Such functions are called \emph{central} or \emph{class} functions.
\end{itemize}
 If the group $\Ga$ consists of
 $M$ elements $\gamma_0,\ldots,\gamma_{M-1}$ and $n_k$ is the number of paths with the 
 ``phase'' $\Phi=\Rep{\gamma_k}$
 at the point of observation $\left(x,t\right)$, then the amplitude at this point
 is $A=\sum\limits_{k=0}^{M-1}n_k\Rep{\gamma_k}\psi$, where $\psi\in\Hspace$. 
The square of the amplitude (i.e., probability after appropriate normalization)
can be written as
\begin{equation}
		\left\langle A\psi|A\psi\right\rangle
		=\sum\limits_{{k=0}}^{{M-1}}n_k^2\left|\psi\right|^2+
		 \sum_{\substack{{\gamma_i,\gamma_k\in\Ga}\\
		 {i<k}}}n_in_k\left\langle\psi\left|\Rep{\gamma_i^{-1}\gamma_k}
		 +\RepH{\gamma_i^{-1}\gamma_k}\right|\psi\right\rangle\enspace,
\end{equation}
or, after collecting like terms, as
\begin{equation}
		\left\langle A\psi|A\psi\right\rangle
		==\sum\limits_{k=0}^{M-1}
N_k\left(n_0,\cdots,n_{M-1}\right)\left\langle\psi\left|\Rep{\gamma_k}
		 +\RepH{\gamma_k}\right|\psi\right\rangle\enspace,		 
\end{equation}
where $N_k\left(n_0,\cdots,n_{M-1}\right)$ are quadratic polynomials with integer coefficients
and arguments. Thus, algebraic integers are sufficient for all our computations  except for 
normalization of probabilities requiring the quotient field of the ring $\A$.  
\subsection{Simple Model Inspired by Free Particle} 
In quantum mechanics -- as is clear from the \emph{never vanishing} expression $\exp\left(\frac{i}{\hbar}S\right)$ for the path 
amplitude -- transitions from one to any other state are possible in principle.
However, we shall consider computationally more tractable models with restricted sets of possible transitions.
\par
Let us consider quantization of a free particle moving in one dimension. 
Such a particle is described
by the Lagrangian $L = \frac{\textstyle{m\dot{x}^2}}{\textstyle{2}}.$ 
Keeping only transitions to the closest points in the discretized space
we come to the following rule for the one-time-step transition amplitudes
\begin{center}
\includegraphics[width=0.20\textwidth]{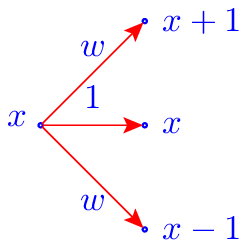}\hspace*{20pt}
\raisebox{34pt}{\begin{tabular}{l}
$e^{\textstyle{\frac{i}{\hbar}\frac{m\left\{(x+1)-x\right\}^2}{2}}}
=e^{\textstyle{i\frac{m}{2\hbar}}}$\\[10pt]
$e^{\textstyle{\frac{i}{\hbar}\frac{m\left(x-x\right)^2}{2}}}\hspace*{17pt}=~1$\\[10pt]
$e^{\textstyle{\frac{i}{\hbar}\frac{m\left\{(x-1)-x\right\}^2}{2}}}=e^{\textstyle{i\frac{m}{2\hbar}}}.$
\end{tabular}}
\end{center}
That is, we have evolution rule as an $\U(1)$-valued function $R$ defined on pairs of points (graph edges).
Symbolically: 
\begin{eqnarray}
	R\left(x\rightarrow{}x\right)&=&1~~\in~~\U(1)\enspace,\nonumber\\
	R\left(x\rightarrow{}x-1\right)=R\left(x\rightarrow{}x+1\right)
	&=&w=e^{\textstyle{i\frac{m}{2\hbar}}}~~\in~~\U(1)\enspace.
	\label{rfree}
\end{eqnarray}
Now let us assume that $w$ in (\ref{rfree}) is an element of some representation of a finite group: 
$w=\Rep{\alpha},~\alpha\in\Ga=\set{\gamma_0=1,\ldots,\gamma_{M-1}}$.
Rearranging \emph{multinomial coefficients} --- \emph{trinomial} in this concrete case --- it is not difficult 
to write the sum amplitude over all paths of the form 
$\left(0,0\right)\longrightarrow\left(x,t\right)$
\begin{equation}
A_x^t\left(w\right)=\sum\limits_{\tau=0}^t\frac{\tau!}{\left(\frac{\tau-x}{2}\right)!\left(\frac{\tau+x}{2}\right)!}
	\times
	\frac{t!}{\tau!\left(t-\tau\right)!}~w^{\tau}\enspace.
\end{equation}
Note that $x$ must lie in the limits determined by $t$: $x\in\left[-t,t\right].$
\par
\begin{figure}[!h]
\centering
\includegraphics[width=0.95\textwidth]{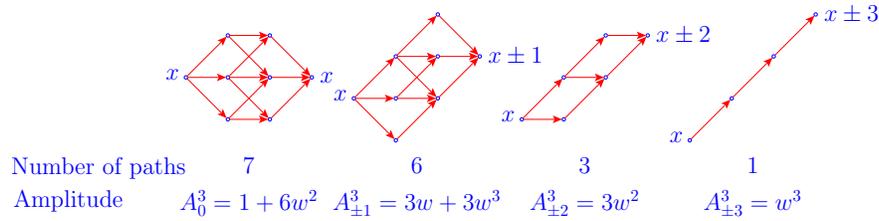}
\caption{Amplitudes for all possible paths in three time steps.}
	\label{Ampl3T}
\end{figure}

One of the most expressive peculiarities of quantum-mechanical behavior is the 
\emph{destructive interference} --- cancellation of non-zero amplitudes attached
to different paths converging to the same point. 
By construction, the sum of amplitudes in our model is a function $A(w)$ depending on distribution 
of sources of the particles, their initial phases,  gauge fields acting along the paths,
restrictions -- like, e.g., ``slits'' -- imposed on possible paths, etc.
In the case of one-dimensional representation the function $A(w)$ is a polynomial with algebraic 
integer coefficients and $w$ is a root of unity. Thus, the condition for destructive interference
can be expressed by the system of polynomial equations: $A(w)=0$ and $w^M=1$.
For concreteness let us consider the cyclic group $\Ga=\Cgr{M}=\set{\gamma_0,\cdots,\gamma_k,\cdots,\gamma_{M-1}}$. 
Any of its $M$ irreducible representations takes the form $\Rep{\gamma_k}=w^k$, 
where $w$ is one of the $M$th roots of unity.
For simplicity let $w$ be the \emph{primitive root}: $w=\e^{2\pi{}i/M}.$
Fig. \ref{Ampl3T} shows all possible transitions from the point $x$ 
in three time steps with their amplitudes.
\par
We see that the polynomial $A^3_{\pm1}=3w+3w^3=3w\left(w^2+1\right)$ contains the
\emph{cyclotomic polynomial} $\Phi_4(w)=w^2+1$ as a factor. The smallest group associated to $\Phi_4(w)$ --- and hence providing the destructive interference --- is $\Cgr{4}$.
Thus, $\Cgr{4}$ is the natural quantizing group for the model under consideration.
\begin{figure}[!h]
\centering
\includegraphics[width=0.95\textwidth]{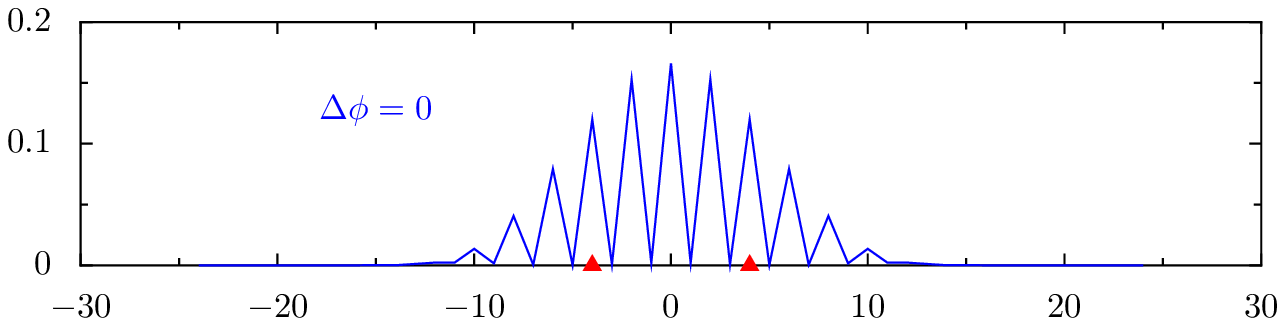}\\
\includegraphics[width=0.95\textwidth]{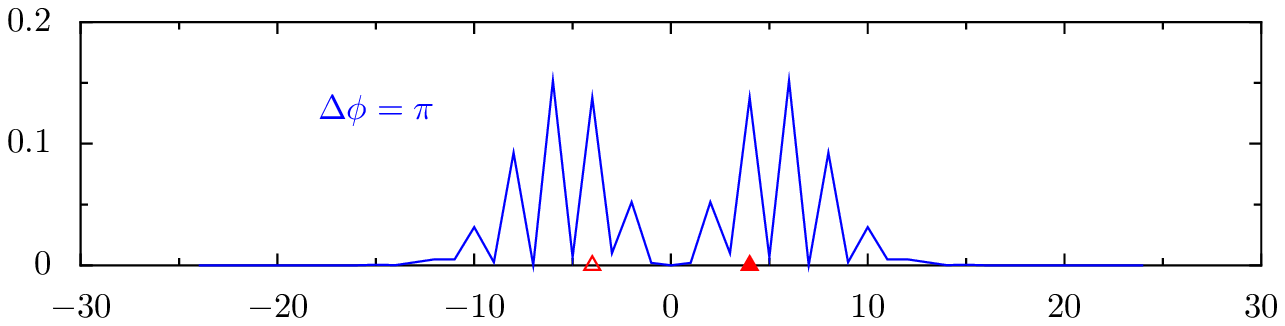}
\caption{Group $\Cgr{4}$. Interference from two sources. 
Number of time steps $T=20$. Source positions are -4 and 4. Phase 
differences $\Delta\phi=\phi_4-\phi_{-4}$ between
sources are $0$ and $\pi$.}
	\label{Interf2}
\end{figure}
\par
Fig. \ref{Interf2} shows interference patterns --- normalized squared amplitudes 
(``probabilities'') ---  from two sources placed in the positions $x=-4$ and $x=4$ 
for 20 time steps. The upper and lower graphs show interference pattern when sources
are in the same ($\Delta\phi=0$) and in the opposite 
($\Delta\phi=\pi$) phases, respectively.
\subsection{Generalization: Local Quantum Model on Regular Graph}
The above model --- with quantum transitions allowed only within the neighborhood of a vertex
of a 1-dimensional lattice --- can easily be generalized to arbitrary regular graph.
Our definition 
of \emph{local quantum model on $k$-valent graph} uncludes the following:
\begin{enumerate}
\item
 \emph{Space} $X=\set{x_1,\cdots,x_N}$ is a $k-$valent graph.
\item
 \emph{Set of local transitions} $E_i=\set{e_{0,i}, e_{1,i},\cdots,e_{k,i}}$ 
 is the set of $k$ adjacent to the vertex $x_i$ edges 	$e_{m,i}=\left(x_i\rightarrow{}x_{m,i}\right)$
 completed by the edge	$e_{0,i}=\left(x_i\rightarrow{}x_i\right)$.
\item
 We assume that the \emph{space symmetry} group $\G=\mathrm{Aut}\left(\X\right)$ acts transitively
 on the set $\set{E_1,\cdots,E_N}$.
\item
 $\G_i=\mathrm{Stab}_{\G}\left(x_i\right)\leq\G$ is the \emph{stabilizer} of $x_i$
	($g\in\G_i$ means $x_ig=x_i$).
\item
 $\Omega_i=\set{\omega_{0,i},\omega_{1,i},\cdots,\omega_{h,i}}$	
	is the \emph{set of orbits} of $G_i$ on $E_i$.
\item
 \emph{Quantizing group} $\Ga$ is a finite group: $\Ga=\set{\gamma_0,\cdots,\gamma_{M-1}}$.
\item
 \emph{Evolution rule} $R$ is a function on $E_i$ with values in some 
 representation $\Rep{\Ga}$. The rule $R$ prescribes $\Rep{\Ga}$-weights 
	to the one-time-step transitions from $x_i$ to elements of the neighborhood of $x_i$.
	From the symmetry considerations $R$ must be a function on orbits from
	$\Omega_i$, i.e., $R\rbra{e_{m,i}g}=R\rbra{e_{m,i}}$ for $g\in\G_i$.
\end{enumerate}
To illustrate these constructions, let us consider the local quantum model on the graph 
of \emph{buckyball}. 
The incarnations of this 3-valent graph include in particular:
\par
-- the \emph{Caley graph of the icosahedral group} $\Alt{5}$ (in mathematics); 
\par
-- the \emph{molecule $C_{60}$} (in carbon chemistry).\\
Here the space $X=\set{x_1,\cdots,x_{60}}$ has the shape
\raisebox{-0.04\textwidth}{\includegraphics[width=0.11\textwidth]{FullereneC60CASTR}}
and its symmetry group is $\G=\mathrm{Aut}\left(\X\right)=\Cgr{2}\times\Alt{5}$.
The set of local transitions takes the form $E_i=\set{e_{0,i},~ e_{1,i},~ e_{2,i},~ e_{3,i}}$, 
where~ $e_{0,i}=\rbra{x_i\rightarrow{}x_i}$,~
 $e_{1,i}=\rbra{x_i\rightarrow{}x_{1,i}}$,\\
 $e_{2,i}=\rbra{x_i\rightarrow{}x_{2,i}}$,~
$e_{3,i}=\rbra{x_i\rightarrow{}x_{3,i}}$
 ~in accordance with~~~~ 
\raisebox{-0.08\textwidth}
{\includegraphics[width=0.2\textwidth]{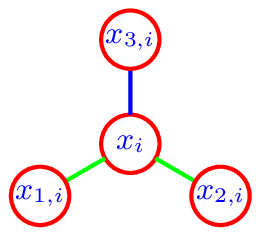}}.\\
The stabilizer of $x_i$ is $\G_i=\mathrm{Stab}_{\G}\left(x_i\right)=\Cgr{2}$. 
The set of orbits of $\G_i$ on $E_i$ contains 3 orbits:
 $\Omega_i=\set{\omega_{0,i}=\set{e_{0,i}}, \omega_{1,i}
=\set{e_{1,i}, e_{2,i}}, \omega_{2,i}=\set{e_{3,i}}}$, i.e.,
the stabilizer does not move the edges $\rbra{x_i\rightarrow{}x_i}$ and
 $\rbra{x_i\rightarrow{}x_{3,i}}$ and swaps $\rbra{x_i\rightarrow{}x_{1,i}}$ and
 $\rbra{x_i\rightarrow{}x_{2,i}}.$ This asymmetry results from different roles the edges
 play in the structure of the buckyball:  $\rbra{x_i\rightarrow{}x_{1,i}}$ and
 $\rbra{x_i\rightarrow{}x_{2,i}}$ are edges of a pentagon adjacent to
 $x_i$ , whereas $\rbra{x_i\rightarrow{}x_{3,i}}$
 separates two hexagons; in the carbon molecule $C_{60}$ the edge $\rbra{x_i\rightarrow{}x_{3,i}}$ 
 corresponds to the double bond, whereas others are the single bonds.
 \par
The evolution rule takes the form:
\begin{center} 
\begin{tabular}{l}
$R\rbra{x_i\rightarrow{}x_i}=\Rep{\alpha_0},$\\
$R\rbra{x_i\rightarrow{}x_{1,i}}=R\rbra{x_i\rightarrow{}x_{2,i}}=\Rep{\alpha_1},$\\
$R\rbra{x_i\rightarrow{}x_{3,i}}=\Rep{\alpha_2},$
\end{tabular}
\end{center}
where $\alpha_0,\alpha_1,\alpha_2\in\Ga$. If we take a one-dimensional representation and move 
$\alpha_0$ -- using gauge invariance -- to the identity element of $\Ga$, we see that the rule $R$ depends
on  $v=\Rep{\alpha_1}$ and  $w=\Rep{\alpha_2}$. 
Thus, the amplitudes in the quantum model on the buckyball take the form $A(v, w)$ depending on two roots of unity.
\section{Conclusion}
Extraordinary success of gauge theories in fundamental physics 
suggests that the gauge principle may be useful in theory and 
applications of discrete dynamical systems also. 
Furthermore, discrete and finite background allowing comprehensive study 
-- especially with the help of computer algebra and methods of computational group theory  
-- may lead 
to deeper understanding of the gauge principle itself and its connection with the quantum behavior. 
To study more complicated models we are developing the C program.
\subsection*{Acknowledgments}
The author thanks Laurent Bartholdi, Vladimir Gerdt and Nikolai Vavilov for useful remarks and comments.
This work was supported in part by the grants 07-01-00660 from the Russian Foundation 
for Basic Research and 1027.2008.2  from the Ministry of Education and Science of the 
Russian Federation.


\begin{thebibliography}{99}
\bibitem{RaifStrau}
O'Raifeartaigh L., Straumann N. 
Gauge theory: Historical Origins and Some Modern Developments. 
\textit{Reviews of Modern Physics,} \textbf{72}, No. 1 (2000) 1--23
\bibitem{Oeckl}
Oeckl R.
\textit{Discrete Gauge Theory (From Lattices to TQPT)}. 
Imperial College Press, London (2005)
\bibitem{Kirillov}
Kirillov A.A. \textit{Elements of the Theory of Representations.}
 Springer-Verlag, Berlin-New York (1976)
\bibitem{Feynman}
Feynman R.P., Hibbs A.R. \textit{Quantum Mechanics and Path Integrals.} McGraw-Hill
(1965) 
\bibitem{Serre}
Serre J.-P. \textit{Linear Representations of Finite Groups.} Springer-Verlag (1977) 
\end{thebibliography}
\end{document}